\documentclass[a4paper,12pt]{article}
\usepackage{typearea}
\typearea{15}
\usepackage{mymacros}
\usepackage{mathtools}
\usepackage{slashed}
\begin{document}
	\rightline{\baselineskip16pt\rm\vbox to20pt{
			{%\hbox{APCTP-Pre2010-xxx}
				\hbox{OCU-PHYS-480}
				\hbox{AP-GR-145}
			}
			\vss}}%

	\begin{center}
		{\LARGE The exact solutions to an Einstein-Dirac-Maxwell system with Sasakian quasi-Killing spinors on 4D spacetimes}\\
		\bigskip\bigskip
		{\large
			Satsuki Matsuno\footnote{smatsuno43@gmail.com}\\
			Fumihiro Ueno\footnote{ueno-fumihiro-dx@alumni.osaka-u.ac.jp}
		}\\
		\bigskip
		{\it Department of Physics, Graduate School of Science,
			Osaka Metropolitan University\\
			3-3-138 Sugimoto, Sumiyoshi, Osaka 558-8585, Japan}
	\end{center}
	
	% \maketitle	
	
	\begin{abstract}
		%There is a global cosmic magnetic field in the Universe.
		Exact solutions to an Einstein-Maxwell(E-D-M) system with an electric current are simple models of global cosmic magnetic phenomena in the universe.
		In this paper, we consider an E-D-M system with two chiral spinors $ \psi^\pm$ coupled with a gauge field $ A$ and an electromagnetic field $F=dA$.
		We construct a family of exact solutions to the E-D-M system on four-dimensional static Sasakian spacetimes.
		The electromagnetic field in the solutions is a contact magnetic field, and the chiral spinors are induced from Sasakian quasi-Killing spinors.
		We find closed universe models whose weak energy condition holds and open universe models whose weak energy condition is violated.
		In this study, we introduce the Sasakian frame, which is used to prospectively discuss the properties of Sasakian quasi-Killing spinors. 
		A simple description of Sasakian quasi-Killing spinors of a certain type is obtained.
	\end{abstract}
	
	\tableofcontents
	
	\section{Introduction}
	In the real universe, there is a somewhat strong global magnetic field\cite{Kulsrud 2008}.
	It is known that there is a strong magnetic field, then there may be an electric current.
	Such a system is composed of Einstein gravity, an electromagnetic field, and matter generating the current.
	Most of the systems are very complicated, and it is often difficult to find exact solutions without simplification.
	
	So, it is interesting to constructing exact solutions to an Einstein-Maxwell system with an electric current.
	Many researchers have constructed exact solutions to Einstein-Maxwell systems with charged perfect fluids\cite{exact solutions}.
	
	Ishihara and Matsuno (one of the present authors) constructed a family of exact solutions to an Einstein-Maxwell system with a charged dust fluid on a static Sasakian spacetime\cite{Ishihara Matsuno 2020,Ishihara Matsuno 2022.2}.
	The solutions are derived from matter consisting of a contact magnetic field and charged dust fluids flowing along Reeb orbits.
	However, as fluid is a composite object, it is expected that similar situations can be created by more fundamental matter, for example spinors or scalar fields.
	
	In \cite{Dzhunushaliev}, on the four-dimensional static Einstein spacetime, a family of exact solutions to Dirac-Maxwell system is constructed.
	The spinor of the solutions generates an electric current which flows along fibers of the Hopf fibration of $S^3$.
	Inspired by this work, we generalize $S^3$ to three-dimensional Sasakian manifolds and construct exact solutions to an Einstein-Dirac-Maxwell (E-D-M) system by considering spinors that generate an electric current along Reeb orbits.
	
	In this paper, we consider an E-D-M system consisting of two chiral spinors $ \psi^\pm$ coupled with a gauge field $ A$ and the electromagnetic field $F=dA$.
	The action functional of the theory is given by
	$$
	S=\int_M\left( \mathcal{R}-\frac{1}{4}||F||^2 +  \Re \lb \overline{\psi^+}(\slashed{D}^+-m^+)\psi^+ \rb  + \Re \lb \overline{\psi^-} (\slashed{D}^--m^-)\psi^- \rb -2 \Lambda\right)\vol_g.
	$$
	
	Our strategy is to use a contact magnetic field and Sasakian quasi-Killing spinors for constructing exact solutions to the system.
	A contact magnetic field is a magnetic field whose gauge field is proportional to a contact form on contact manifolds, which was introduced in \cite{Cabrerizo 2009}.
	The motion of charged particles in contact magnetic fields and other geometrical properties have been studied in \cite{Cabrerizo 2009,Romaniuc 2015,Inoguchi Munteanu 2022} (and references therein).
	However, few studies have investigated the relationship between a contact magnetic field and Einstein gravity.
	A Sasakian quasi-Killing spinor is a generalization of a Killing spinor in Sasakian geometry, which was introduced by Friedrich and Kim\cite{Kim Friedrich 2000}.
	Killing spinors play an important role in supergravity and have been actively studied\cite{Friedrich Thomas 2003,Semmelmann 1998,Fujii 1886,Felipe 2003} (and references therein).
	On the other hand, Sasakian quasi-Killing spinors seem to have few applications in mathematical physics.
	It is expected that using Sasakian quasi-Killing spinors rather than Killing spinors prescribes a rich solution space to Einstein gravity since the number of Sasakian quasi-Killing spinors is larger than that of Killing spinors on Sasakian manifolds.
	
	We construct a family of exact solutions to the E-D-M system on static Sasakian spacetimes, which are direct product manifolds of time and a Sasakian 3-manifold.
	The matter fields in the solutions are a contact magnetic field and chiral spinors made from Sasakian quasi-Killing spinors.
	These solutions describe a situation where an electric current flows uniformly along one direction in space, generating a global cosmic magnetic field.
	There are closed universe models homeomorphic to $\mathbb{R}\times S^3$ whose weak energy condition holds and open universe models homeomorphic to $\mathbb{R}\times\mathbb{R}^3$ whose weak energy condition is violated.
	
	The organization of this paper is as follows.
	At the \hyperref[subsc:relusts]{end} of this section, we summarize the main results of this study.
	In Section \ref{sc:The Einstein-Dirac-Maxwell system} we describe our system in detail.
	We also review some facts about Sasakian geometry and Sasakian quasi-Killing spinors, and explain how to construct a spinor on a four-dimensional static Sasakian spacetime from spinors on a three-dimensional Sasakian manifold.
	Section \ref{sc:Solutions to the Einstein-Dirac-Maxwell system} is devoted to solving our system.
	First, we introduce the Sasakian frame, which is used to prospectively discuss the properties of Sasakian quasi-Killing spinors and show some propositions.
	Next, after fully determining the solutions of the Einstein-Dirac system in our settings, we construct a family of exact solutions to our E-D-M system.
	Some of the calculations are described in Appendix \ref{sc:Calculations in the section 3}.
	Finally, in Section \ref{sc:Conclusions and discussions}, the results of this paper are briefly summarized and the energy condition of spinors and possible future research are discussed.

	\subsection*{Summary of the results}\label{subsc:relusts}
	We summarize our results in \pref{sc:results}.
	Let $(N,\phi,\xi,\eta,h)$ be a three-dimensional Sasakian space form of $\phi$-sectional curvature $H$, and let $(M,g)$ be a four-dimensional static spacetime defined by $$
	M=\mathbb{R}\times N,\ g=-dt^2+r^2h,\ r\in\mathbb{R}_{>0}.
	$$
	We consider the following matter fields on $(M,g)$.
	The electromagnetic field is $F=dA$ whose gauge field is given by $A=B\eta,\ B\in\mathbb{R}$.
	The massive spinor fields are $\psi^\pm=e^{\sqrt{-1}Et}\phi_R^\pm\oplus\phi_L^\pm$ with charges $\pm q$ and masses $m^\pm>0$ respectively, where $\phi_P^\pm,\ P=R,L$ are Sasakian quasi-Killing spinors of type $(a,b)$.
	Suppose $(M,g),F$ and $\psi^\pm$ satisfy the E-D-M system, that is,
	\begin{align}
		&\Ric -\frac{1}{2}\mathcal{R}g + \Lambda g = T \coloneqq T^{\rm em}+T^{\rm spin},\\
		&(\slashed{D}^\pm - m^\pm)\psi^\pm = 0,\\
		&*d*F  = {}^\flat \lb J_q(\psi^+) + J_{-q}(\psi^-) \rb,
	\end{align}
	where $\Lambda\in\mathbb{R}$ is cosmological constant, $\slashed{D}^\pm$ are Dirac operators of spin-c connection with respect to charge $\pm q$, and $J_{\pm q}(\psi^\pm)$ are the Dirac current of $\psi^\pm$.
	Then the solutions of the system for $H\ne-5$ are given by
	\begin{align}
		(a,b)&=\left(\frac{1}{2},\frac{H-1}{4}\right),\\
		||\phi^+_R||_N^2=||\phi^-_R||_N^2&=- \frac{B \left(4 B q  + 4 E r + H + 5\right)}{q r^{3}(4Bq+H+5)},\\
		||\phi^+_L||_N^2=||\phi^-_L||_N^2&=\frac{B \left(4 B q - 4 E r + H + 5\right)}{q r^{3} \left(4 B q + H + 5\right)},\\
		(m^+)^2&= - \frac{\left(H + 4  B q  + 5- 4rE\right) (H+4 B q   + 5+ 4 E r)}{16 r^{2}},\\
		(m^-)^2&=-\frac{\left(H - 4  B q+5 -4r E \right) \left(H-4 B q +5 + 4 E r \right)}{16 r^{2} },\\
		B^2&=  \frac{\left(H - 1\right) \left(L + 1\right)r^2}{2(H + 5)},\\
		q^2&= \frac{(H+5)\left(H - 1\right) \left(L + 1\right)}{ 8r^2\left(H - 3 L + 2\right)^2 },\\
		E^2&= - \frac{ \left(H^{2} - 2 H L + 6 H - 4 L + 11\right) \left(H^{2} - 2 H L + 8 H - 16 L + 9\right)}{16 r^2 \left(H - 3 L + 2\right)^2},
	\end{align}
	where $L=\Lambda r^2$, and the relation of the Sasakian space form $N$ and the range of $H,L$ and $E$ are given in the following table:
	\begin{table}[H]
		\caption{The ranges of $ H, L $ and $ E $}
		\centering
		\begin{tabular}{|c|c|c|c|c|}
			\hline
			& Range of $H$ & Range of $L$ & Sign of $E$ & $ N $ \\
			\hline\hline
			(i) & $-8<H<-5$  & $h(H)<L<\infty$ & Negative & $ \widetilde{{\rm SL}(2,\bR)} $ \\
			\hline
			(ii) & $-5<H<-4$  & $k(H)<L<-1$ & Negative & $ \widetilde{{\rm SL}(2,\bR)} $ \\
			\hline
			(iii) & $1<H$  & $h(H)<L<k(H)$ & Positive & $ S^3 $ \\
			\hline
		\end{tabular}
	\end{table}
	\noindent where $ k(H)=\frac{H+2}{2},\  h(H)=\frac{H^2+8H+9}{2(H+8)}$.

	\section{The Einstein-Dirac-Maxwell system}\label{sc:The Einstein-Dirac-Maxwell system}
	First, we describe our E-D-M system. 
	Let $ (M, g) $ be a four-dimensional spacetime. We consider two spinors $ \psi^\pm $ coupled with a $ U(1) $ gauge field $ A $ and an electromagnetic field $ F=dA $. 
	In this paper, we solve Euler-Lagrange equations of the following Lagrangian density of the E-D-M system
	\begin{equation}\label{eq:Lagrangian density of E-D-M}
		\mathcal{L} = \mathcal{R}-\frac{1}{4}||F||^2 + \lb \psi^+, (\slashed{D}^+-m^+)\psi^+ \rb_M + \lb \psi^-, (\slashed{D}^--m^-)\psi^- \rb_M -2 \Lambda,
	\end{equation} 
	where $ \Lambda $ is a cosmological constant, $ \scR $ is the scalar curvature, $ m^{\pm} > 0$ are masses of $ \psi^{\pm} $, $ \slashed{D}^{\pm} $ are Dirac operators with respect to a spin-c connections $\nabla^\pm q$ for charges $\pm q$, and $\lb \psi_1, \psi_2 \rb_M = \Re(\overline{\psi_1}\psi_2) $.
	The Euler-Lagrange equations are Einstein, Dirac and Maxwell equations
	\begin{align}
		&\Ric -\frac{1}{2}\mathcal{R}g + \Lambda g = T \coloneqq T^{\rm em}+T^{\rm spin}, \label{eq:Einstein}\\
		&(\slashed{D}^\pm - m^\pm)\psi^\pm = 0, \label{eq:dirac}\\
		&*d*F  = {}^\flat \lb J_q(\psi^+) + J_{-q}(\psi^-) \rb \label{eq:maxwell},
	\end{align}
	where the Dirac currents of $ \psi^\pm $ are defined by $ J_{\pm q}(\psi^\pm) \coloneqq \pm q(\overline{\psi^\pm} \gamma^\mu \psi^\pm) X_\mu$ using a global orthonormal frame $ \{X_\mu\}_{\mu=0}^3 $ of $ (M,g) .$ 
	The energy-momentum tensors of $ \psi^\pm $ are given by
	\begin{equation}\label{eq:energytensor of spinors}
		T^{\psi^\pm}(X,Y)=\frac{1}{4}\lb \psi^\pm,\sqrt{-1}X\cdot\nabla_{Y}^{\pm q}\psi^\pm+\sqrt{-1}Y\cdot\nabla_{X}^{\pm q}\psi^\pm \rb_M, ( X, Y \in \Gamma(TM)).
	\end{equation}
	Then the total energy-momentum tensor of spinors is written as 
	$T^{\rm spin} = T^{\psi^+} + T^{\psi^-}.  $
	The energy-momentum tensor of the electromagnetic field $ F $ is given by 
	\begin{equation}\label{eq:energytensor of electromagnetic}
		T^{\rm em}(X, Y) = g^*(X \lrcorner F  , Y \lrcorner F ) -\frac{1}{4}||F||^2g(X, Y), (X, Y \in \Gamma(TM)).
	\end{equation}

	Secondary, we review some basic facts about Sasakian geometry. 
	The definition of Sasakian manifolds is as follows.
	\begin{definition}\label{df:Sasaki mfd}
		A contact metric manifold is a quintuple $(M, \phi, \xi, \eta, g)$ consisting of an odd dimensional manifold $ M $, a $(1,1)$-type tensor field $\phi$, a vector field $\xi$, a one-form $\eta$ and a Riemannian metric $g$, which satisfies the following conditions.
		\begin{align}
			\phi^2 &= -\id + \eta\otimes\xi,\\
			\eta(\xi)&=1, \\
			g(\phi X, \phi Y) &= g(X, Y) - \eta(X)\eta(Y),\\
			g(X, \phi Y) &= d\eta(X, Y).
		\end{align}
		The quadruple $ (\phi,\xi, \eta, g) $ is called a contact metric structure of $ M$.
		The one-form $ \eta $ and the vector field $ \xi $ are called the contact form and the Reeb vector field of the contact metric structure.
		A contact manifold $(M, \phi, \xi,\eta, g)$ is called a Sasakian manifold if the Riemann curvature tensor of $ (M, g) $ satisfies $ R(X,Y)\xi = \eta(Y)X-\eta(X)Y $ for arbitrary vector fields $ X, Y \in \Gamma(TM) $. 
	\end{definition}
	It is known that a three-dimensional Riemannian manifold $ (M,g) $ is a Sasakian manifold if and only if $ M $ admits a unit Killing vector field $ \xi $ and constant sectional curvature for any plane section including $\xi$. 
	For any vector field $ X $ orthogonal to $ \xi $, the sectional curvature of the form $ H(X) \coloneqq K(X, \phi(X)) $ is called a $ \phi- $sectional curvature.
	A complete simply connected Sasakian manifold which has constant $\phi-$sectional curvature is called a Sasakian space form.
	
	Third, we introduce spacetimes which we investigated. Let $ (N, h) $ be a three-dimensional simply connected Sasakian space form, and $ (M, g) $ be a four-dimensional static spacetime defined by $ M = \bR \times N $ and $ g = -dt^2 + r^2h ,$ where $ t $ is a coordinate of $\bR$ and $ r $ is a positive constant.
	Because Sasakian space forms are all Lie groups, thus $ (N, h) $ has a unique spin structure, therefore, $ (M, g) $ also has a unique spin structure.
	
	Finally, we introduce an ansatz of matter fields. We define a gauge field $ A=B\eta, (B \in \bR) $ using the contact form $ \eta $ on $ N. $ Then the electromagnetic field $ F $ is defined as $ F = dA, $ and since it is pure magnetic, it is called a contact magnetic field.
	Our two chiral spinors of spinor bundle $ S_M $ over $ (M,g) $ are derived from Sasakian quasi-Killing spinors of spinor bundle $ S_N $ over $ (N, h)$.
	A Sasakian quasi-Killing spinor is defined by Friedrich and Kim\cite{Kim Friedrich 2000}.
	%which satisfies a certain differential equation.
	\begin{definition}\label{df:Sasakian-quasi-Killing}
		A Sasakian quasi-Killing spinor of type $ (a, b) $ is a non-trivial spinor field $ \phi $ which satisfies the following equation
		\begin{equation}\label{eq:Sasakian-quasi-Killing}
			\nabla^{S_N}_X \phi = a X \cdot \phi + b\eta(X)\xi\cdot \phi,
		\end{equation}
		where, $ a,b$ are real numbers and $ \nabla^{S_N} $ is the spin connection induced from the Levi-Civita connection of $ (N,h) $.
	\end{definition}
	% 	\begin{comment}
		% 		Spinors $ \phi_R$ and $\phi_L$ of $ S_N $ induce two spinors $ \iota_P(\phi_P) \in \Gamma(S^P_M) \subset \Gamma(S^R_M \oplus S^L_M)=\Gamma(S_M)$ on $ S_M (P=R, L)$ 
		% 		using Riemannian submersion $ \pi \colon M \to N$, 
		% 		where $ \iota_P $ is the inclusion map.% we explain in the following(soon, immediately).
		% 		In generally, globally spliting staionary spacetime $ (M,g) $ might not reduce the structure group from $ Spin(1,3) $ to $ Spin(3) $ of the spinor bundle whereas there is an inclusion of bundle map $ \iota_P \colon \Gamma(S_N) \hookrightarrow \Gamma(S_M) $. In addition, if $ (M,g) $ is a globally spliting static spacetime, there exists a bundle isomorphism $ \pi^*S_N \simeq S^P_M. $
		% 	\end{comment}
	
	The spinor bundle $ S_M $ has a chiral decomposition $ S_M = S^R_M \oplus S^L_M$.
	We have bundle isomorphisms $S_M^P\simeq \pi^\ast S_N\ (P=R,L)$ where $\pi:M=\mathbb{R}\times N\to N$ is a projection map onto the second factor.
	Via these isomorphisms, we obtain two inclusion maps $\iota_P \colon \Gamma(S_N) \hookrightarrow \Gamma(S^R_M \oplus S^L_M) = \Gamma(S_M)$, that is, for $\phi\in\Gamma(S_N) $, we define $ \iota_P(\phi) \in \Gamma(S^P_M)$ as
	\begin{align}\label{eq:inclution}
		&\iota_R(\phi)=
		\begin{pmatrix}
			\phi \\ 0
		\end{pmatrix} \in \Gamma(S^R_M),\ 
		\iota_L(\phi)=
		\begin{pmatrix}
			0 \\ \phi
		\end{pmatrix} \in \Gamma(S^L_M).
	\end{align}
	For four Sasakian quasi-Killing spinors $ \phi^\pm_P(P=R, L) $ of $ S_N $, we can consider two chiral spinors $ \psi^{\pm} $ of $ S_M $
	\begin{equation}\label{eq:two chiral spinors}
		\psi^{\pm} \coloneqq e^{\sqrt{-1}Et}\lb \iota_R(\phi^\pm_R) + \iota_L(\phi^\pm_L)  \rb 
		= e^{\sqrt{-1}Et}\lb \phi^\pm_R \oplus \phi^\pm_L  \rb 
		= e^{\sqrt{-1}Et} 
		\begin{pmatrix}
			\phi^\pm_R \\ \phi^\pm_L
		\end{pmatrix}, 
		(E \in \bR).
	\end{equation}
	
	To describe an interaction between charged spinors and gauge fields, we would like to introduce the spin-c connection.
	Let $L$ be a trivial complex line bundle and let $A=B\eta,\ B\in\mathbb{R}$.
	We define two connection forms $\pm \sqrt{-1}qA$ on $L$ as gauge potentials.
	We identify the section  of the spinor bundle $ S_M$ with a section of a trivial spin-c bundle $ \Sigma = S_M \otimes L $.  Let $ \nabla^{\pm q} = \nabla^{S_M} \pm \sqrt{-1}q A $ be two connections on $ \Sigma $ for charges $ \pm q(q > 0).$
	Since the spin connection on $ S_M $ splits as $ \nabla^{S_M} = \nabla^{S^R_M} \oplus \nabla^{S^L_M},$ we obtain a connection decomposition on $ \Sigma = \Sigma^R \otimes \Sigma^L(\Sigma^P = S_M^P \otimes L )$ as
	\begin{equation}\label{eq: gauged spin connection of chiral spinor}
		\nabla^{\pm q} = (\nabla^{S^R_M} \pm \sqrt{-1}qA) \oplus (\nabla^{S^L_M} \pm \sqrt{-1}qA). 
	\end{equation}   
	We use Pauli matrices $ \sigma^i(i=1, 2, 3) $ as a basis of Lie algebra $ \sqrt{-1}\mathfrak{so}(3) $, i.e.
	\begin{equation}
		\sigma^1
		=\begin{pmatrix}
			0 & 1 \\
			1& 0
		\end{pmatrix},
		\sigma^2
		=\begin{pmatrix}
			0 & -\sqrt{-1} \\
			\sqrt{-1}& 0
		\end{pmatrix},
		\sigma^3
		=\begin{pmatrix}
			1 & 0 \\
			0& -1
		\end{pmatrix}.
	\end{equation}
	The gamma matrices in the chiral representation are written as
	\begin{equation}\label{eq:Gamma matrics}
		\gamma^\mu = 
		\begin{pmatrix}
			0 & \sigma^\mu \\  \sigmabar^\mu & 0
		\end{pmatrix} (\mu=0,1,2,3),
	\end{equation}
	where $ \sigma_0=I_2 $ and $ (\sigmabar^0, \sigmabar^1, \sigmabar^2, \sigmabar^3)=(\sigma^0, -\sigma^1, -\sigma^2, -\sigma^3) $. Let $ \slashed{D}^\pm = \sqrt{-1}\sum_{\mu}\gamma^\mu \nabla^{\pm q}_{X_\mu} $ be Dirac operators on $ \Sigma $ using a global orthonormal frame $ \{X_\mu\}_{\mu = 0}^3 $ of $ (M,g).$ The decomposition of Dirac operators
	\begin{equation}
		\slashed{D}^\pm =
		\begin{pmatrix}
			0 & \slashed{D}^\pm_L \\
			\slashed{D}^\pm_R & 0
		\end{pmatrix} \\ 
	\end{equation}
	holds, 
	where $ \slashed{D}^\pm_R = \sqrt{-1}\sum_{\mu}\sigmabar^\mu(\nabla^{S^R_M}_{X_\mu}\pm \sqrt{-1}q A(X_\mu)) $ and  $ \slashed{D}^\pm_L = \sqrt{-1}\sum_{\mu}\sigma^\mu(\nabla^{S^L_M}_{X_\mu} \pm \sqrt{-1}q A(X_\mu))  $.
	
	Let $ \la \phi_1, \phi_2 \ra_N $ denote a Hermitian metric on $ S_N $ for $ \phi_i \in \Gamma(S_N), (i=1,2) $ and $ \la \psi_1, \psi_2 \ra_M = \overline{\psi_1}\psi_2 $ denote an invariant bilinear form on $ S_M $ for $ \psi_i \in \Gamma(S_M)(i=1,2), $ where $ \overline{\psi_1}\coloneqq \psi_1^\dagger \gamma^0 $ is the Dirac conjugate of $ \psi_1 \in \Gamma(S_M). $
	We also use brackets $  \lb \bullet,  \bullet \rb_N = \Re{\la \bullet, \bullet \ra_N}$ and $ \lb \bullet , \bullet \rb_M = \Re{\la \bullet, \bullet \ra_M}$ and use the same symbols for $ \Sigma$.

	\section{Solutions to the Einstein-Dirac-Maxwell system}\label{sc:Solutions to the Einstein-Dirac-Maxwell system}
	In this section, we solve our E-D-M system.
	We describe all solutions to our system except in the case $H=-5$.
	% We discuss two case, that is massive spinors case and massless spinors case.
	
	%%%%%%%%%%%%%%%%%%%%%%%%%%%%%%%%%%%%%%%%%%%%%%%%%%%%%%%%%%%%%%%%%%%%%%%%%%%%
	\subsection{Preparation for solving the E-D-M system}
	At first, we derive some basic formulae and propositions.
	three-dimensional Sasakian space forms are homogeneous and there are three kinds.
	They are decided by $ \phi$-sectional curvature $ H$.
	A Sasakian space form is in  Bianchi type IX if $ H>-3$, type II if $ H=-3$, or type VIII if $ H<-3$.
	They are diffeomorphic to $ {\rm S}^3,{\rm Nil}_3,\widetilde{{\rm SL}(2,\mathbb{R})}$ respectively.
	
	There is a useful global orthonormal frame on a Sasakian space form.
	On a three-dimensional Sasakian space form $ (N, \phi, \xi, \eta, h)$, there exists a global orthonormal frame $ \{e_1=\xi,e_2,e_3\}$ which satisfies 
	\begin{align}
		[e_1,e_2]=-\frac{H+3}{2}e_3,\ [e_1,e_3]=\frac{H+3}{2}e_2,\ [e_2,e_3]=-2e_1,\label{eq:sasakian frame}
	\end{align}
	where $ H $ is a $ \phi$-sectional curvature (see Appendix \ref{sc:sasakian frame}).
	We call this frame a Sasakian frame.
	The Ricci tensor of $ N$ is given by 
	\begin{align}
		{\rm Ric}_N={\rm diag}(2,H+1,H+1),
	\end{align}
	with respect to a Sasakian frame $ \{e_1,e_2,e_3\}$.
	
	We can define a global orthonormal frame $ \{X_0,X_1,X_2,X_3\}$ on $ (M,g)$, where $ M=\mathbb{R}\times N$ and $ g=-dt^2+r^2h$.
	That is, let $ \{e_1,e_2,e_3\}$ be a Sasakian frame on $ (N,h)$ and put $ X_0=\partial_t,\ X_i=\frac{1}{r}e_i,\ (i=1,2,3)$.
	We also call this frame a Sasakian frame.
	Moreover the Ricci tensor of $ (M,g)$ is given by 
	\begin{align}
		{\rm Ric}_M={\rm diag}\left(0,\frac{2}{r},\frac{H+1}{r},\frac{H+1}{r}\right),
	\end{align}
	with respect to a Sasakian frame $ \{X_0,X_1,X_2,X_3\}$.
	
	Let $ \nabla^N,\nabla^M$ denote the Levi-Civita connections of $ N,M$ respectively.
	The Levi-Civita connection $ \nabla^N$  is given by 
	\begin{align}
		\nabla^N_{e_i}e_j=-\sum_{k=1}^3c_i\epsilon_{ijk}e_k,\ (i,j\in \{1, 2, 3\}),
	\end{align}
	for a Sasakian frame $ \{e_1,e_2,e_3\}$, where $ c_1=\frac{H+1}{2},\ c_2=c_3=1$.
	
	We denote the co-frame of $ \{e_1,e_2,e_3\}$ by $ \{\omega^1=\eta,\omega^2,\omega^3\}$, where $\eta$ is the contact form of $(N,h)$.
	Hereafter we identify a $p$-form $ \alpha \in \Omega^p(N)$ and $ \pi^\ast \alpha \in \Omega^p(M)$.

	%%%%%%%%%%%%%%%%%%%%%%%%%%%%%%%%%%%%%%%%%%%%%%%%%%%%%%%%%%%%%%%%%%%%%%%%%%%%
	\subsubsection{Formulae and Propositions for spinor fields}\label{subsubsec:Formulae and Propositions for spinor fields}
	Here we give the connection form of the spinor bundle $S_M$ and some basic propositions.
	We use a matrix representation of Clifford algebra  $ \bC l_{3}  $ written as
	\begin{equation}\label{eq:Clifford-three-rep}
		e_i  \mapsto \sqrt{-1}\sigma_i \eqqcolon s_i \ (i=1,2,3),
	\end{equation} % \Gamma(TM) \ni X_\mu  \mapsto \sigma_\mu \in Cl(TM)(\mu=0,1,2,3) 
	where $ \{e_1, e_2, e_3\} $ is a Sasakian frame on $ N $.
	\begin{lemma}
		The spin connection form $ \omega^{S_N}$ of the spinor bundle $ S_N$ is given by 
		\begin{align}
			\omega^{S_N}=\frac{1}{2}\sum_{k=1}^3s_k\omega^k+\frac{H-1}{4}s_1\eta,\label{eq:spinor connection of N}
		\end{align}
		with respect to a Sasakian frame $\{e_1,e_2,e_3\}$.
		Furthermore, the spin connection form $ \omega^{S_M}$ of the spinor bundle $ S_M$ is given by \begin{align}
			\omega^{S_M}=\frac{1}{r}\omega^{S_N}\oplus \frac{1}{r}\omega^{S_N}.
		\end{align}
		Then, for $ \phi\in\Gamma(S_N)$, we have $ \nabla^{S_M}\iota_P(\phi)=\iota_P(\nabla^{S_N}\phi)$.
	\end{lemma}
	\begin{proof}
		The first claim follows from the simple calculation
		\begin{align}
			\omega^{S_N}&=\frac{1}{4}\sum_{i,j,k}h(\nabla_{e_i}e_j,e_k)\omega^is_js_k=-\frac{1}{4}\sum_{i,j,k}h(\nabla_{e_i}e_j,e_k)\omega^i\epsilon_{jkl}s_l\nonumber\\
			&=\frac{1}{2}c_3\omega^3 s_3 +\frac{1}{2}c_2\omega^2 s_2+\frac{1}{2}c_1\omega^1 s_1\nonumber\\
			&=\frac{1}{2}\sum_k s_k\omega^k+\frac{H-1}{4}s_1\eta,\nonumber
		\end{align}
		and the second claim is derived by
		\begin{align}
			\omega^{S_M}&=\frac{1}{4}\sum_{\mu,\nu,\lambda}g(\nabla_{X_\mu}X_\nu,X_\lambda)\Tilde{\omega}^\mu\gamma^\nu\gamma^\lambda
			=\frac{1}{4r}\sum_{i,j,k}h(\nabla_{e_i}e_j,e_k)\tilde\omega^i\gamma^j\gamma^k\\
			&=\frac{1}{4r}\sum_{i,j,k}h(\nabla_{e_i}e_j,e_k)\tilde\omega^i(s_js_k\oplus s_js_k)
			=\frac{1}{r}\omega^{S_N}\oplus \frac{1}{r}\omega^{S_N},
		\end{align}
		where $\{\tilde\omega^\mu\}$ is the co-frame of a Sasakian frame $\{X_\mu\}$.
	\end{proof}
	
	% We identify a vector field $ X\in\Gamma(TN)$ and its horizonal lift with respect to $ \pi$ such that $ [\partial_t,X]=0$
	
	Using a Sasakian frame $ \{e_1, e_2, e_3\} $, an invariant Hermitian metric is given by $ \langle\phi_1,\phi_2\rangle_N = \phi_1^\dagger\phi_2$ and
	the Dirac current of $ \phi\in\Gamma(S_N)$ is given by
	$ J_N(\phi)\coloneqq-\sum_i\sqrt{-1}\langle\phi,s_i\phi\rangle_Ne_i$.
	We denote $\sum_iJ_N(\phi)_i$ for an i-th component of $J_N(\phi)$.
	
	Friedrich and Kim have classified Sasakian quasi-Killing spinors on three-dimensional Sasakian space forms as follows\cite{Kim Friedrich 2000}.
	
	\begin{proposition}\label{pr:Friedrich classification}
		On three-dimensional Sasakian space forms with $\phi$-sectional curvature $H$, there exists a Sasakian quasi-Killing spinor of type $ \left(\frac{1}{2},\frac{H-1}{4}\right)$.
		Moreover, if $ H>-3$ then there exists a Sasakian quasi-Killing spinor of type $ \left(\frac{1\mp \sqrt{3+H}}{2},-\frac{2\mp\sqrt{3+H}}{2}\right)$.
		In particular, if $ H=-\frac{3-\sqrt{5}}{2}$ then there exists a Sasakian quasi-Killing spinor of type $ \left(-\frac{3+\sqrt{5}}{4},\frac{5+\sqrt{5}}{4}\right)$, and if $ H=-\frac{3+\sqrt{5}}{2}$ then there exists a Sasakian quasi-Killing spinor of type $ \left(-\frac{3-\sqrt{5}}{4},\frac{5-\sqrt{5}}{4}\right)$.
	\end{proposition}
	
	The connection form \eqref{eq:spinor connection of N} enables us to obtain a simple description of a Sasakian quasi-Killing spinor of type $\left(\frac{1}{2}, \frac{H-1}{4}\right)$ as follows.
	\begin{proposition}\label{pr:const saskian quasi-Killing spinor}
		A spinor $ \phi\in\Gamma(S_N)$ is a Sasakian quasi-Killing spinor of type $ \left(\frac{1}{2}, \frac{H-1}{4}\right)$ if and only if $ \phi=\begin{pmatrix}\alpha \\ \beta\end{pmatrix},\ \alpha,\beta\in\mathbb{C}$ with respect to a Sasakian frame $ \{e_1,e_2,e_3\}$.
	\end{proposition}

	We decompose the spinor bundle $ S_N$ as well as Friedrich.
	The spinor bundle $ S_N$ splits into the orthogonal direct sum $ S_N=\Sigma_0\oplus\Sigma_1$ with $ \xi|_{\Sigma_0}=-\sqrt{-1}\ {\rm Id},\ \xi|_{\Sigma_1}=\sqrt{-1}\ {\rm Id}$.
	Sections $\chi_0\in\Gamma(\Sigma_0)$ and $\chi_1\in\Gamma(\Sigma_1)$ are given by $\chi_0=f_0\begin{pmatrix}1\\ -1\end{pmatrix}$ and $\chi_1=f_1\begin{pmatrix}1\\ 1\end{pmatrix}$ for $ f_0, f_1\in C^\infty(N)$ 
	with respect to a Sasakian frame since we have $\xi\cdot=s_1=\begin{pmatrix} 0 & \sqrt{-1} \\ \sqrt{-1} & 0\end{pmatrix}$ by \eqref{eq:Clifford-three-rep}. 
	In particular, a Sasakian quasi-Killing spinor of type $\left(\frac{1}{2},\frac{H-1}{4}\right)$ is in $\Gamma(\Sigma_0\oplus\Sigma_1)$ from \pref{pr:const saskian quasi-Killing spinor}.

	%%%%%%%%%%%%%%%%%%%%%%%%%%%%%%%%%%%%%%%%%%%%%%%%%%%%%%%%%%%%%%%%%%%%%%%%%%%%
	\subsubsection{Formulae of the electromagnetic field}
	Here we explicitly write down the Maxwell equation \eqref{eq:maxwell} and the energy-momentum tensor of the electromagnetic field \eqref{eq:energytensor of electromagnetic}.
	
	Since $ A=B\eta $ and $F=dA,$ we have
	\begin{align}
		\ast d\ast F&=\ast d\ast(2B\omega^2\wedge\omega^3)
		=\ast d\left(\frac{2B}{r}\omega^0\wedge \omega^1\right)\nonumber\\
		&=\ast\left(-\frac{4B}{r}\omega^0\wedge\omega^2\wedge\omega^3\right)
		=\frac{4B}{r^2}\omega^1,
	\end{align}
	it follows that the Maxwell equation \eqref{eq:maxwell} is given by
	\begin{align}
		\frac{4B}{r^3}X_1=J_q(\psi^+)+J_{-q}(\psi^-).\label{eq:maxwell 2}
	\end{align}
	
	By simple calculation, we have the energy-momentum tensor of $F$ as follows
	\begin{align}
		T^{\rm em}=\frac{2B^2}{r^4}{\rm diag}(1,-1,1,1),\label{eq:energytensor of electromagnetic 2}
	\end{align}
	with respect to a Sasakian frame $\{X_0,X_1,X_2,X_3\}$.

	%%%%%%%%%%%%%%%%%%%%%%%%%%%%%%%%%%%%%%%%%%%%%%%%%%%%%%%%%%%%%%%%%%%%%%%%%%%%
	\subsubsection{Basic propositions of the Dirac-Maxwell system}
	In our settings, Maxwell equation \eqref{eq:maxwell} uniquely decides the type of Sasakian quasi-Killing spinors.
	% We recall our settings.
	\begin{comment}
		Let $ \phi^\pm_P \ (P=R,L)$ be four Sasakian quasi-Killing spinors of type $ (a,b)$ of $ S_N$, $ \psi^\pm=e^{iEt}\begin{pmatrix}\phi^\pm_R \\ \phi^\pm_L\end{pmatrix}\ (E\in\mathbb{R})$ be two chiral spinors with charge $ \pm q$ respectively,  $ A=B\eta$ be a gauge field, where $ \eta$ is the contact form of $ N$, and $F=dA$ be the electromagnetic field.
	\end{comment}
	First, we consider the derivative of the components of the Dirac current of a Sasakian quasi-Killing spinor as follows.
	\begin{lemma}\label{lm:derivative of current}
		Let $N$ be a three-dimensional Sasakian space form, and let $\{e_1,e_2,e_3\}$ be a Sasaian frame.
		For a Sasakian quasi-Killing spinor $\phi$ of type $ (a,b)$, the derivative of the scalar function $J_N(\phi)_i$ is given by
		\begin{align}
			e_j(J_N(\phi)_i) =\sum_k(2a\epsilon_{jik}+2b\eta(e_j) \epsilon_{i1k} -c_j\epsilon_{jik})J_N(\phi)_k, \label{eq:derivative of current}
		\end{align}
		especially $e_3(J_N(\phi)_2)=(-2a+1)J_N(\phi)_1$ holds.
	\end{lemma}
	\begin{proof}
		\begin{align}
			e_j(J_N(\phi)_i)&=\sqrt{-1}\langle\nabla_j\phi,s_i\phi\rangle_N+\sqrt{-1}\langle\phi,(\nabla_je_i)\phi\rangle_N+\sqrt{-1}\langle\phi,s_i\nabla_j\phi\rangle_N\nonumber\\
			&=a\sqrt{-1}\langle s_j\phi,s_i\phi\rangle_N+a\sqrt{-1}\langle\phi,s_is_j\phi\rangle_N)+ b\eta(e_j)\sqrt{-1}\langle s_1\phi,s_i\phi\rangle_N\nonumber\\
			&+b\sqrt{-1}\eta(e_j)\langle\phi,s_is_1\phi\rangle_N+\sqrt{-1}\langle\phi,(\nabla_{e_j}e_i)\phi\rangle_N\nonumber\\
			&=a\sqrt{-1}\langle\phi,-s_js_i+s_is_j\phi\rangle_N+b\sqrt{-1}\eta(e_j)\langle\phi,-s_1s_i+s_is_1\phi\rangle_N\nonumber\\
			&+\sqrt{-1}\langle\phi,(\nabla_{e_j}e_i)\phi\rangle_N\nonumber\\
			&=2a\sum_k\epsilon_{jik}J_N(\phi)_k+2b\eta(e_j) \sum_k\epsilon_{i1k} J_N(\phi)_k-c_j\sum_k\epsilon_{jik}J_N(\phi)_k\nonumber\\
			&(\text{we used} \ s_is_j = -\delta_{ij}-\sum_k\epsilon_{ijk}s_k {\rm\ and\ }\nabla_{e_i}e_j=-\sum_kc_i\epsilon_{ijk}e_k)\nonumber\\
			&=\sum_k(2a\epsilon_{jik}+2b\eta(e_j) \epsilon_{i1k} -c_j\epsilon_{jik})J_N(\phi)_k
		\end{align}
		
	\end{proof}
	
	Then the following proposition holds.
	\begin{proposition}\label{pr:maxwell decides type}
		Let $ \psi^\pm \in \Gamma(\Sigma)$ be two chiral spinors with charge $ \pm q $ defined by \eqref{eq:two chiral spinors}. 
		If $ \psi^\pm$ and $A$ satisfy the Maxwell equation \eqref{eq:maxwell}, then $ (a,b)=\left(\frac{1}{2}, \frac{H-1}{4}\right)$.
	\end{proposition}
	\begin{proof}
		Since \eqref{eq:maxwell 2}, the current $ J_{q}(\psi^+)+J_{-q}(\psi^-)$ is proportional to $\xi$, so we have
		\begin{align}
			J_1&\coloneqq J_{q}(\psi^+)_1+J_{-q}(\psi^-)_1=q(-J_N(\phi^+_R)_1+J_N(\phi^+_L)_1+J_N(\phi^-_R)_1-J_N(\phi^-_L)_1)\ne0,\\
			J_2&\coloneqq J_{q}(\psi^+)_2+J_{-q}(\psi^-)_2=q(-J_N(\phi^+_R)_2+J_N(\phi^+_L)_2+J_N(\phi^-_R)_2-J_N(\phi^-_L)_2)=0,
		\end{align}
		and furthermore $e_3(J_2)=0$ must holds.
		Since \pref{lm:derivative of current}, we have
		\begin{align}
			e_3(J_2)&=q(-2a+1)(-J_N(\phi^+_R)_1+J_N(\phi^+_L)_1+J_N(\phi^-_R)_1-J_N(\phi^-_L)_1)\nonumber\\
			&=q(-2a+1)J_1.
		\end{align}
		therefore we have $ a=\frac{1}{2}$.
		By \pref{pr:Friedrich classification}, the classification result by Friedrich and Kim, $ b=\frac{H-1}{4}$ holds.
	\end{proof}
	
	\begin{remark}
		From \pref{eq:derivative of current} we have 
		$ e_1(J_N(\phi)_2) =(2a-2b -c_1)J_N(\phi)_3,\ 
		e_1(J_N(\phi)_3) =(-2a+2b +c_1)J_N(\phi)_2,\ 
		e_2(J_N(\phi)_3) =(2a-1)J_N(\phi)_1,\ 
		e_3(J_N(\phi)_2) =(-2a +1)J_N(\phi)_1,\ 
		e_2(J_N(\phi)_2) =e_3(J_N(\phi)_3) =0$.
		Thus, in the case $(a,b)=\left(\frac{1}{2},\frac{H-1}{4}\right)$, if $J_N(\phi)_2=J_N(\phi)_3=0$ at an arbitrary point then $J_N(\phi)_2=J_N(\phi)_3=0$ on the whole of $N$.
		Then $J_N(\phi)$ is a Killing vector field because \pref{eq:derivative of current} implies
		$ (\nabla_{e_j}J_N(\phi))_i=\sum_k(2a\epsilon_{jik}+2b\eta(e_j) \epsilon_{i1k})J_N(\phi)_k$. 
		Thus, on a three-dimensional Sasakian space form with $H\ne 1$, a Sasakian quasi-Killing spinor $\phi$ of type $\left(\frac{1}{2},\frac{H-1}{4}\right)$ with $J_N(\phi)_2=J_N(\phi)_3=0$ is an example of a non-Killing spinor whose Dirac current is a Killing vector field.
		If $H=1$, it is a Killing spinor.
	\end{remark}
	
	As we intend to show in \pref{pr:D-M in massive case}, the Dirac equation leads to that $ \phi_P^\pm$ are in $ \Gamma(\Sigma_0)$ or $ \Gamma(\Sigma_1)$.
	Given this fact, a more concrete representation of Maxwell's equations is as follows.
	\begin{proposition}
		Let $ \psi^\pm \in \Gamma(\Sigma)$ be two chiral spinors with charge $ \pm q $ defined by \eqref{eq:two chiral spinors}. 
		If Sasakian quasi-Killing spinors $\phi^\pm_P,\ P\in\{R,L\}$ of type $\left(\frac{1}{2}, \frac{H-1}{4}\right)$ are in $\Gamma(\Sigma_{\ell_\pm}),\ \ell_\pm\in\{0,1\}$,
		then the Maxwell equation \eqref{eq:maxwell} is given by
		\begin{align}
			&||\phi^+_R||_N^2+||\phi^+_L||_N^2-||\phi^-_R||_N^2-||\phi^-_L||_N^2=0,\label{eq:maxwell 0-compo}\\
			&(-1)^{\ell_+}||\phi^+_R||_N^2-(-1)^{\ell_+}||\phi^+_L||_N^2-(-1)^{\ell_-}||\phi^-_R||_N^2+(-1)^{\ell_-}||\phi^-_L||_N^2=\frac{4B}{qr^3}\label{eq:maxwell 1-compo}
		\end{align}
	\end{proposition}
	\begin{proof}
		By the assumptions, we have
		\begin{align}
			\phi_R^\pm=\alpha_R^\pm\begin{pmatrix}1 \\ (-1)^{1+\ell_\pm}\end{pmatrix},\ 
			\phi_L^\pm=\alpha_L^\pm\begin{pmatrix}1 \\ (-1)^{1+\ell_\pm}\end{pmatrix},\ \alpha_R^\pm,\alpha_L^\pm\in\mathbb{C}.
		\end{align}
		Then we obtain
		\begin{align}
			J_q(\psi^+)+J_{-q}(\psi^-)&=q \lb \langle\psi^+,\gamma^0\psi^+\rangle_M-\langle\psi^-,\gamma^0\psi^-\rangle_M\rb X_0
			+q\lb \langle\psi^+,\gamma^1\psi^+\rangle_M-\langle\psi^-,\gamma^1\psi^-\rangle_M \rb X_1\nonumber\\
			&=2q\lb |\alpha_R^+|^2+|\alpha_L^+|^2-|\alpha_R^-|^2-|\alpha_L^-|^2 \rb X_0\nonumber\\
			&+2q\lb (-1)^{\ell_+}(|\alpha_R^+|^2-|\alpha_L^+|^2)-(-1)^{\ell_-}(|\alpha_R^-|^2-|\alpha_L^-|^2) \rb X_1\nonumber\\
			&=q\lb ||\phi^+_R||_N^2+||\phi^+_L||_N^2-||\phi^-_R||_N^2-||\phi^-_L||_N^2 \rb X_0\nonumber\\
			&+q \lb (-1)^{\ell_+}||\phi^+_R||_N^2-(-1)^{\ell_+}||\phi^+_L||_N^2-(-1)^{\ell_-}||\phi^-_R||_N^2+(-1)^{\ell_-}||\phi^-_L||_N^2 \rb X_1,\nonumber
		\end{align}
		which implies the claim.
	\end{proof}
	
	Next we write down the Dirac equation \eqref{eq:dirac}.
	\begin{proposition}\label{pr:dirac2}
		Let $ \psi^\pm \in \Gamma(\Sigma)$ be two chiral spinors with charge $ \pm q $ defined by \eqref{eq:two chiral spinors}. 
		The Dirac equations $(\slashed{D}^\pm-m^\pm)\psi^\pm=0$ are equivalent to the following linear equations
		\begin{align}
			D^\pm_R\phi^\pm_R &= m^\pm\phi^\pm_L,\\
			D^\pm_L\phi^\pm_L &= m^\pm\phi^\pm_R,
		\end{align}
		where $ D^\pm_R=\left(-E+\frac{3a+b}{r}\right) I_2\mp\frac{qB}{r}\sigma_1$ and $
		D^\pm_L=\left( -E-\frac{3a+b}{r}\right)I_2\mp\frac{qB}{r}\sigma_1 .$ 
	\end{proposition}
	\begin{proof}
		We can verify it by simple calculation,
		\begin{align}
			\slashed{D}^\pm\psi^\pm&=
			\begin{pmatrix}
				0 & \sqrt{-1}\sigma^\mu\lb \nabla^{S^L_M}_{X_\mu} \pm \sqrt{-1}q A(X_\mu)\rb \\
				\sqrt{-1}\sigmabar^\mu \lb \nabla^{S^R_M}_{X_\mu}\pm \sqrt{-1}q A(X_\mu) \rb & 0
			\end{pmatrix}\psi^\pm \nonumber\\
			&=\begin{pmatrix}0 & -E-(3a+b)/r\mp\frac{qB}{r}\sigma_1 \\ -E+(3a+b)/r\pm\frac{qB}{r}\sigma_1 & 0\end{pmatrix}\psi^\pm.
		\end{align}
	\end{proof}

	%%%%%%%%%%%%%%%%%%%%%%%%%%%%%%%%%%%%%%%%%%%%%%%%%%%%%%%%%%%%%%%%%%%%%%%%%%%%
	\subsection{Solutions to the E-D-M system}\label{sc:results}
	In this section, we begin by fully determining the solutions to our D-M system.
	It then becomes clear that there are two cases whether $ H = -5$ or not and that the solutions to the D-M system are quite different for each case.
	Then, after calculating the energy-momentum tensor of the spinors, we solve to our E-D-M system for the case of $ H\ne-5$.
	\begin{proposition}\label{pr:D-M in massive case}
		Let $ \psi^\pm \in \Gamma(\Sigma)$ be two chiral spinors with charge $ \pm q $ defined by \eqref{eq:two chiral spinors}.
		Suppose that $ \psi^\pm$ and $A$ satisfy the D-M system then the following hold.
		
		%{\rm (I)}
		\begin{enumerate}[{\normalfont (I)}]
			\item In the case of $ H\ne-5$, we have $ \phi^\pm_P\in\Gamma(\Sigma_{\ell_\pm}),\ \ell_\pm\in\{0,1\}$.
			Especially, 
			\begin{enumerate}[{\normalfont (i)}]
				\item if $ (\ell_+,\ell_-)=(1,0)$, we have
				\begin{align}
					||\phi^+_R||_N^2=||\phi^-_R||_N^2&=- \frac{B \left(4 B q  + 4 E r + H + 5\right)}{q r^{3}(4Bq+H+5)}\label{eq:phi_R in massive case 1},\\
					||\phi^+_L||_N^2=||\phi^-_L||_N^2&=\frac{B \left(4 B q - 4 E r + H + 5\right)}{q r^{3} \left(4 B q + H + 5\right)},\label{eq:phi_L in massive case 1}
				\end{align}
				\item if $ (\ell_+,\ell_-)=(0,0)$, we have
				\begin{align}
					||\phi^+_R||_N^2&= \frac{4 B q - 4 E r - H - 5}{4 q^{2} r^{3}},\ 
					||\phi^-_R||_N^2= - \frac{4 B q + 4 E r + H + 5}{4 q^{2} r^{3}},\\
					||\phi^+_L||_N^2&=- \frac{4 B q + 4 E r - H - 5}{4 q^{2} r^{3}},\ 
					||\phi^-_L||_N^2= \frac{4 B q - 4 E r + H + 5}{4 q^{2} r^{3}},
				\end{align}
				\item if $ (\ell_+,\ell_-)=(0,1)$, we have the same results as in the case $ ({\rm i}) $ without the sign of $q$.\\
				\item if $ (\ell_+,\ell_-)=(1,1)$, we have the same results as in the case $ ({\rm ii}) $ without the sign of $q$.
				
			\end{enumerate}
			\item In the case of $ H=-5$, we have $ \phi^\pm_p\in\Gamma(\Sigma_0\oplus\Sigma_1)$ and $ \phi^\pm_L=\frac{1}{m}\begin{pmatrix}-E & \frac{B q}{r}\\\frac{B q}{r} & -E\end{pmatrix}\phi^\pm_R$.
		\end{enumerate}
	\end{proposition}
	\begin{proof}
		Since $\psi^\pm $ and $A$ satisfy the Maxwell equation and \pref{pr:maxwell decides type}, then $a=\frac{1}{2},b=\frac{H-1}{4}$ and $ \phi^\pm_p\in\Gamma(\Sigma_0\oplus\Sigma_1)$ holds.
		Thus, by \pref{pr:dirac2}, the Dirac equations for $\psi^\pm$ are given by
		\begin{align}
			&D^\pm_R\phi^\pm_R = m^\pm \phi^\pm_L,\ D^\pm_L\phi^\pm_L = m^\pm \phi^\pm_R,\label{eq:dirac decomp}\\
			&D^\pm_R=\left(-E+\frac{H+5}{4r}\right) I_2\mp\frac{qB}{r}\sigma_1,\label{eq:dirac decomp r}\\
			&D^\pm_L=-\left( E+\frac{H+5}{4r}\right)I_2\mp\frac{qB}{r}\sigma_1,\label{eq:dirac decomp l}
		\end{align}
		then we obtain
		\begin{align}
			&D^\pm_RD^\pm_L\phi_R^\pm=(m^\pm)^2\phi_R^\pm,\label{eq:eigen equation R}\\
			&D^\pm_RD^\pm_L\phi_L^\pm=(m^\pm)^2\phi_L^\pm,\label{eq:eigen equation L}\\
			&D^\pm_RD^\pm_L=D^\pm_LD^\pm_R=\begin{pmatrix}- \frac{16B^{2} q^{2} r^{2} + \left(H + 4E r + 5\right) \left(H - 4 E r + 5\right)}{16r^{2}} & \mp\frac{B q \left( H  + 5\right)}{2 r}\\ \mp\frac{B q \left( H + 5\right)}{2 r} & - \frac{16B^{2} q^{2} r^{2} + \left(H + 4E r + 5\right) \left(H - 4 E r + 5\right)}{16r^{2}}\end{pmatrix}\nonumber\\
			&\ \ \ \ \ \ \ \ \ \eqqcolon \begin{pmatrix}\alpha & \beta^\pm \\ \beta^\pm & \alpha\end{pmatrix}
		\end{align}
		
		(I) In the case of $H\ne-5$, that is $\beta^\pm \ne0$, an eigenvector of $D^\pm_RD^\pm_L$ corresponding to the eigenvalue $\alpha+\beta^\pm$ is a section of $\Sigma_1$ and that corresponding to the eigenvalue $\alpha-\beta^\pm$ is a section of $\Sigma_0$.
		Then, by \eqref{eq:eigen equation R} and \eqref{eq:eigen equation L}, $ \phi_P^\pm  $ belong to the eigenspace $ \Sigma_0$ or $ \Sigma_1 $ respectively and $ \phi_R^+, \phi_L^+$ ( resp. $ \phi_R^-, \phi_L^-$) are in the same eigenspace.
		Therefore we have $\phi_R^+, \phi_L^+ \in\Gamma(\Sigma_{\ell_+}),\ \phi_R^-, \phi_L^-\in\Gamma(\Sigma_{\ell_-}),\ \ell_\pm\in\{0,1\}$.
		Since \pref{pr:const saskian quasi-Killing spinor}, we have
		\begin{align}
			\phi_R^\pm=\alpha_R^\pm\begin{pmatrix}1 \\ (-1)^{1+\ell_\pm}\end{pmatrix},\ 
			\phi_L^\pm=\alpha_L^\pm\begin{pmatrix}1 \\ (-1)^{1+\ell_\pm}\end{pmatrix},\ \alpha_R^\pm,\alpha_L^\pm\in\mathbb{C}.
		\end{align}
		
		Then \eqref{eq:dirac decomp} implies
		\begin{align}
			m^\pm=\frac{\alpha^\pm_R}{\alpha^\pm_L}\left(\frac{H - 4 E r + 5+(-1)^{1+\ell_\pm} 4Bq}{4 r}\right)
			=-\frac{\alpha^\pm_L}{\alpha^\pm_R}\left(\frac{H + 4 E r + 5+(-1)^{1+\ell_\pm} 4Bq}{4 r}\right).
		\end{align}
		If $\ell_\pm=0$, we have
		\begin{align}
			\frac{||\phi_R^\pm||^2_N}{||\phi_L^\pm||^2_N}&=\left(\frac{\alpha_R}{\alpha_L}\right)^2=  -\frac{4 B q  - 4 E r - H -5}{4 B q  +4 E r - H - 5},\label{eq:phi ratio-0} \\
			(m^\pm)^2&=-\frac{\left(H - 4  B q+5 -4r E \right) \left(H-4 B q +5 + 4 E r \right)}{16 r^{2} }(=\alpha-\beta^\pm), 
		\end{align}
		and if $\ell_\pm=1,$ we have
		\begin{align}
			\frac{||\phi_R^\pm||^2_N}{||\phi_L^\pm||^2_N}&=\left(\frac{\alpha_R}{\alpha_L}\right)^2= - \frac{4 B q  + 4 E r + H + 5}{4 B q  - 4 E r + H + 5},\label{eq:phi ratio-1}\\
			(m^\pm)^2&= - \frac{\left(H + 4  B q  + 5- 4rE\right) (H+4 B q   + 5+ 4 E r)}{16 r^{2}}(=\alpha+\beta^\pm).
		\end{align}
		Then we can write $||\phi^\pm_R||_N^2,\ ||\phi^\pm_L||_N^2$ as functions of $B,q,E,r, $ and $H$ as follows.
		Since \eqref{eq:maxwell 0-compo}, we have
		\begin{align}
			\frac{||\phi^+_L||^2_N}{||\phi^-_L||^2_N}=\frac{(||\phi^-_R||^2_N/||\phi^-_L||^2_N)+1}{(||\phi^+_R||^2_N/||\phi^+_L||^2_N)+1},
		\end{align}
		furthermore, since \eqref{eq:maxwell 1-compo}, we obtain
		\begin{align}
			\frac{1}{||\phi^-_L||_N^2}=\frac{qr^3}{4B}\left((-1)^{\ell_+}\left(\frac{||\phi^+_R||_N^2}{||\phi^+_L||_N^2}-1\right)\frac{||\phi^+_L||_N^2}{||\phi^-_L||_N^2}-(-1)^{\ell_-}\left(\frac{||\phi^-_R||_N^2}{||\phi^-_L||_N^2}-1\right)\right).
		\end{align}
		Since $||\phi_R^\pm||^2_N/||\phi_L^\pm||^2_N,\ ||\phi_R^\pm||^2_N/||\phi_L^\pm||^2_N$ are given by \eqref{eq:phi ratio-0} and \eqref{eq:phi ratio-1}, we can decide $||\phi^-_L||_N^2$ via the above two equations. The others $||\phi^+_L||^2_N,||\phi_R^\pm||^2_N$ are given by $||\phi^+_L||^2_N=(||\phi^+_L||^2_N/||\phi^-_L||^2_N)||\phi^-_L||^2_N$ and $||\phi_R^\pm||^2_N=(||\phi_R^\pm||^2_N/||\phi_L^\pm||^2_N)||\phi_L^\pm||^2_N$.
		By simple calculation, our claims are obtained.\\
		
		(II) In the case of $H=-5$, we have
		\begin{align}
			D^\pm_R &= \left(\begin{matrix}-E & B q/r\\B q/r & -E\end{matrix}\right),\ 
			D^\pm_L = \left(\begin{matrix}-E & - B q/r \\- B q/r & -E\end{matrix}\right),\\
			D^\pm_RD^\pm_L&=(E^2-B^2q^2/r^2)I_2,
		\end{align}
		then the claim immediately holds.
	\end{proof}
	
	We calculate the energy-momentum tensor of spinors for our system.
	We shall use the following proposition to calculate the total energy-momentum tensor of spinors in \pref{pr:energy of spinors for 4-cases}.
	
	\begin{proposition}\label{pr:spin-energy-calculation}
		Let $ \phi_R, \phi_L$ be Sasakian quasi-Killing spinors of type $ (a,b) $ of $ S_N$, $ \psi = \psi_R \oplus \psi_L = e^{\sqrt{-1}Et}\iota_R(\phi_R) \oplus e^{\sqrt{-1}Et}\iota_L(\phi_L)  $ be a chiral spinor of $ \Sigma $ with charge $ q \in \bR$, and $ A=B\eta$ be a gauge field with $ B \in \bR $. Then the energy-momentum tensor of $ \psi$ is given by
		%-------------------
		\begin{align}
			T^{\psi}(X_0, X_0) & = \frac{1}{2}E\lb ||\phi_R||^2_N + ||\phi_L||^2_N \rb, \\ \notag
			T^{\psi}(X_0, X_i) & = \frac{1}{4} \lc \lb \frac{a}{r} + E \rb J_N(\phi_R)_i + \lb \frac{b}{r}J_N(\phi_R)_1  + \frac{qB}{r}||\phi_R||^2_N \rb \delta_{1 i} \rc\\ 
			& + \frac{1}{4} \lc \lb \frac{a}{r} - E \rb J_N(\phi_L)_i  + \lb \frac{b}{r}J_N(\phi_L)_1  + \frac{qB}{r}||\phi_L||^2_N \rb \delta_{1 i} \rc ,\\ \notag
			T^{\psi}(X_i, X_j) & = \frac{1}{4}\lc  \frac{2a}{r}||\phi_R||^2_N \delta_{ij} + \frac{2b}{r}\delta_{1i}\delta_{1j} ||\phi_R||^2_N + \frac{qB}{r}\lb J_N(\phi_R)_i\delta_{1j} + J_N(\phi_R)_j\delta_{1i} \rb  \rc \\ 
			& - \frac{1}{4}\lc  \frac{2a}{r}||\phi_L||^2_N \delta_{ij} + \frac{2b}{r}\delta_{1i}\delta_{1j} ||\phi_L||^2_N + \frac{qB}{r}\lb J_N(\phi_L)_i\delta_{1j} + J_N(\phi_L)_j\delta_{1i} \rb  \rc,
		\end{align}
		where $ \{ X_0, X_1, X_2, X_3\} $ is a Sasakian frame and $ i, j \in \{1, 2,3\} $. 
		%--------------------
		
	\end{proposition}
	\begin{proof}
		The spin connection of  the spinor $ \psi \coloneqq e^{\sqrt{-1}Et}\lb \iota_R(\phi_R) + \iota_L(\phi_L)  \rb = \psi_R \oplus \psi_L \in \Gamma(\Sigma)$ is following,  
		\begin{align}\notag
			\lb \nabla^{S^P_M}_{X_\nu}+ \sqrt{-1}q A(X_\nu) \rb \phi_P e^{\sqrt{-1}Et} & = \begin{cases}
				\sqrt{-1}E\phi_P e^{\sqrt{-1}Et} (\nu =0),\\
				\sqrt{-1}\lc \frac{a}{r}\sigma^\nu + \lb \frac{b}{r}\sigma^1 + \frac{qB}{r} \rb \delta_{1\nu} \rc \phi_P e^{\sqrt{-1}Et} (\nu=1,2,3),
			\end{cases}
		\end{align}
		\begin{equation}\notag
			\nabla_{X_\nu}^{+q}\psi = \lb \nabla^{S^R_M}_{X_\mu}+ \sqrt{-1}q A(X_\mu) \rb \phi_R e^{\sqrt{-1}Et} \oplus \lb \nabla^{S^L_M}_{X_\mu}+ \sqrt{-1}q A(X_\mu) \rb \phi_L e^{\sqrt{-1}Et}.
		\end{equation}		
		Since $ \nabla^{S^R_M} = \nabla^{S^L_M}={}^{\pi^*}\nabla^{S_N}, $ and $ \iota_R(\phi_R)=\phi_R \oplus 0, \iota_L(\phi_L)=0 \oplus \phi_L$, it follows,
		\begin{equation}\notag
			\la \psi, \sqrt{-1}X_\mu \cdot \nabla^{+q}_{X_\nu} \psi \ra_M = \sqrt{-1}\lb \phi_R^\dagger \sigmabar_\mu (\nabla^{S^R_M}_{X_\nu} + \sqrt{-1}qA(X_\nu))\phi_R + \phi_L^\dagger \sigma_\mu (\nabla^{S^L_M}_{X_\nu} + \sqrt{-1}qA(X_\nu))\phi_L \rb.
		\end{equation}
		Then these imply that 
		\begin{align}
			\la \psi, \sqrt{-1}X_0 \cdot \nabla^{+q}_{X_0} \psi \ra_M & = E\lb ||\phi_R||^2 + ||\phi_L||^2 \rb, \\
			\la \psi, \sqrt{-1}X_0 \cdot \nabla^{+q}_{X_i} \psi \ra_M & =\phi_R^\dagger \lc \frac{a}{r}\sigma^i + \lb \frac{b}{r}\sigma^1 + \frac{qB}{r} \rb \delta_{1 i} \rc \phi_R+(R \leftrightarrow L), \\ 
			\la \psi, \sqrt{-1}X_i \cdot \nabla^{+q}_{X_0} \psi \ra_M & = E(\phi_R^\dagger \sigma^i \phi_R - \phi_L^\dagger \sigma^i \phi_L), \\ \notag
			\la \psi, \sqrt{-1}X_i \cdot \nabla^{+q}_{X_j} \psi \ra_M & = \frac{a}{r}\phi^\dagger_R\sigma^i\sigma^j\phi_R + \lb \frac{b}{r}\phi_R^\dagger \sigma^i\sigma^1\phi_R + \frac{qB}{r}\phi_R^\dagger \sigma^i \phi_R \rb\delta_{1j}\\
			& - (R \leftrightarrow L).
		\end{align}
		Thus we have
		\begin{align}
			T^{\psi}(X_0, X_0) & = \frac{1}{2}E\lb ||\phi_R||^2 + ||\phi_L||^2 \rb, \\
			T^{\psi}(X_0, X_i) & = \frac{1}{4} \phi_R^\dagger \lc \lb \frac{a}{r} + E \rb\sigma^i + \lb \frac{b}{r}\sigma^1 + \frac{qB}{r} \rb \delta_{1 i} \rc \phi_R\\ 
			& + \frac{1}{4} \phi_L^\dagger \lc \lb \frac{a}{r} - E \rb\sigma^i + \lb \frac{b}{r}\sigma^1 + \frac{qB}{r} \rb \delta_{1 i} \rc \phi_L,\\ \notag
			T^{\psi}(X_i, X_j) & = \frac{1}{4}\lc  \frac{2a}{r}||\phi_R||^2 \delta^{ij} + \frac{2b}{r}\delta^{1i}\delta_{1j} ||\phi_R||^2 + \frac{qB}{r}\lb \phi^\dagger_R \sigma^i \phi_R\delta_{1j} + \phi^\dagger_R \sigma^j \phi_R\delta_{1i} \rb  \rc \\ 
			& - \frac{1}{4}\lc  \frac{2a}{r}||\phi_L||^2 \delta^{ij} + \frac{2b}{r}\delta^{1i}\delta_{1j} ||\phi_L||^2 + \frac{qB}{r}\lb \phi^\dagger_L \sigma^i \phi_L\delta_{1j} + \phi^\dagger_L \sigma^j \phi_L\delta_{1i} \rb  \rc,
		\end{align}
		by using $ \sigma^i\sigma^j + \sigma^j \sigma^i = 2\delta^{ij} $ and $J_N(\phi_P) = \phi_P^\dagger \sigma^i \phi_P e_i.$
	\end{proof}
	
	Let $T^{\rm spin}=T^{\psi_+}+T^{\psi_-}$ be the total energy-momentum tensor of spinors $\psi^\pm$ defined by \eqref{eq:two chiral spinors}. 
	Hereafter, we consider only the case $ H\neq -5.$ 
	
	\begin{proposition}\label{pr:energy of spinors for 4-cases}
		Let $ \psi^\pm \in \Gamma(\Sigma)$ be two chiral spinors with charge $ \pm q $ defined by \eqref{eq:two chiral spinors}. 
		Suppose that $H\ne-5$ and that $ \psi^\pm$ and $A$ satisfy D-M system, which implies $ \phi^\pm_P\in\Gamma(\Sigma_{\ell_\pm}),\ \ell_\pm\in\{0,1\}$.
		Then $T^{\rm spin}$ is given by
		\begin{align}
			&T^{\rm spin}=\begin{pmatrix}
				T^{\rm spin}(X_0,X_0) & T^{\rm spin}(X_0,X_1) & 0 & 0\\
				T^{\rm spin}(X_1,X_0) & T^{\rm spin}(X_1,X_1) & 0 & 0\\
				0 & 0 & T^{\rm spin}(X_2,X_2) & 0\\
				0 & 0 & 0 & T^{\rm spin}(X_3,X_3)
			\end{pmatrix},\\
			&T^{\rm spin}(X_2,X_2)=T^{\rm spin}(X_3,X_3),
		\end{align}
		with respect to a Sasakian frame $\{X_0,X_1,X_2,X_3\}$, and especially we have the following.
		
		\begin{enumerate}[{\normalfont (i)}]
			\item If $ (\ell_+,\ell_-)=(1,0)$, we have
			\begin{align}
				T^{\rm spin}(X_0,X_0)&= - \frac{8 B E^{2}}{q r^{2} \cdot \left(4 B q + H + 5\right)},\ 
				T^{\rm spin}(X_0,X_1)=0,\label{eq:energy00 of spinor in massive case 1}\\ 
				T^{\rm spin}(X_1,X_1)&=- \frac{B \left(4 B q + H + 1\right)}{2 q r^{4}},\label{eq:energy11 of spinor in massive case 1}\\
				T^{\rm spin}(X_2,X_2)&=T^{\rm spin}(X_3,X_3)= - \frac{B}{q r^{4}},\label{eq:energy22 of spinor in massive case 1}
			\end{align}
			\item if $ (\ell_+,\ell_-) = (0,1)$, we have the same results as the case of {\rm(i)} without sign of $ q $,\\
			\item if $ (\ell_+,\ell_-) = (1,1)$, we have
			\begin{align}
				T^{\rm spin}(X_0,X_0)&= - \frac{2 E^{2}}{q^{2} r^{2}},\label{eq:energy00 of spinor in massive case 2}\\ 
				T^{\rm spin}(X_0,X_1)&=-\frac{ E \left(H + 3\right)}{2q^2 r^{3} },\label{eq:energy01 of spinor in massive case 2}\\ 
				T^{\rm spin}(X_1,X_1)&= - \frac{16 B^{2} q^{2} + H^{2} + 6 H + 5}{8 q^{2} r^{4}},\label{eq:energy11 of spinor in massive case 2}\\
				T^{\rm spin}(X_2,X_2)&=T^{\rm spin}(X_3,X_3)= - \frac{H + 5}{4 q^{2} r^{4}},\label{eq:energy22 of spinor in massive case 2}
			\end{align}
			\item if $ (\ell_+,\ell_-) = (0,0)$, we have the same results as the case (iii) without sign of $T^{\rm spin}(X_0,X_1)$.
		\end{enumerate}
	\end{proposition}
	\begin{proof}
		For $ \phi^\pm_P\in\Gamma(\Sigma_{\ell_\pm}),\ \ell_\pm\in\{0,1\}$ we have $J_N(\psi_P^\pm)_2=J_N(\psi_P^\pm)_3=0$, and then \pref{pr:spin-energy-calculation} implies $T^{\rm spin}(X_0,X_2)=T^{\rm spin}(X_0,X_3)=T^{\rm spin}(X_1,X_2)=T^{\rm spin}(X_1,X_3)=T^{\rm spin}(X_2,X_3)=0$.
		Moreover, from \pref{pr:spin-energy-calculation}, other components of $T^{\rm spin}$ are given by
		\begin{align}
			T^{\rm spin}(X_0,X_0)&=\frac{E}{2}(||\phi^+_R||^2_N+||\phi^+_L||^2_N+||\phi^-_R||^2_N+||\phi^-_L||^2_N),\\
			T^{\rm spin}(X_1,X_1)&=\frac{H+1}{8r}(||\phi^+_R||^2_N-||\phi^+_L||^2_N+||\phi^-_R||^2_N-||\phi^-_L||^2_N)\nonumber\\
			&+\frac{qB}{2r}(J_N(\phi^+_R)_1-J_N(\phi^+_L)_1-J_N(\phi^-_R)_1+J_N(\phi^-_L)_1),\\
			T^{\rm spin}(X_2,X_2)&=T^{\rm spin}(X_3,X_3)=\frac{1}{4r}(||\phi^+_R||^2_N-||\phi^+_L||^2_N+||\phi^-_R||^2_N-||\phi^-_L||^2_N),\\
			T^{\rm spin}(X_0,X_1)&=\frac{H+1}{16r}(J_N(\phi^+_R)_1+J_N(\phi^+_L)_1+J_N(\phi^-_R)_1+J_N(\phi^-_L)_1)\nonumber\\
			&+\frac{E}{4}(J_N(\phi^+_R)_1-J_N(\phi^+_L)_1+J_N(\phi^-_R)_1-J_N(\phi^-_L)_1)\nonumber\\
			&+\frac{qB}{4r}(||\phi^+_R||_N^2+||\phi^+_L||_N^2-||\phi^-_R||_N^2-||\phi^-_L||_N^2).
		\end{align}
		Substituting $||\phi_P^\pm||_N^2$ of \pref{pr:D-M in massive case} into the above equations with $J_N(\phi_P^\pm)_1=(-1)^{1+\ell_\pm}||\phi_P^\pm||_N^2$, we obtain our claim by a slightly longer calculation.
	\end{proof}
	
	Now we determine the solutions to our system with $H\ne-5$.
	Since we have \pref{pr:D-M in massive case} and \pref{pr:energy of spinors for 4-cases}, it is sufficient to consider the case $ (\ell_+,\ell_-)=(1,0)$ and $ (\ell_+,\ell_-)=(1,1)$.
	For both cases, the Einstein equations are given by
	\begin{align}
		\frac{2+H}{r^2}-\Lambda&=\frac{2B^2}{r^4}+T^{\rm spin}_{00},\label{eq:ein00}\\
		T^{\rm spin}_{01}&=0,\label{eq:ein01}\\
		-\frac{H}{r^2}+\Lambda&=-\frac{2B^2}{r^4}+T^{\rm spin}_{11},\label{eq:ein11}\\
		-\frac{1}{r^2}+\Lambda&=\frac{2B^2}{r^4}+T^{\rm spin}_{22}\label{eq:ein22},
	\end{align}
	where \eqref{eq:ein01} trivially holds in the case of $ (\ell_+,\ell_-)=(1,0)$.
	
	%%%%%%%%%%%%%%%%%%%%%%%%%%%%%%%%%%%%%%%%%%%%%%%%%%%%%%%%%%%%%%%%%%%%%%%%%%%%
	\subsubsection{The case $ (\ell_+,\ell_-)=(1,0)$}\label{subsec:massive case 1}
	The solutions to the Einstein equations \eqref{eq:ein00},\eqref{eq:ein11},\eqref{eq:ein22} with \eqref{eq:energy00 of spinor in massive case 1},\eqref{eq:energy11 of spinor in massive case 1},\eqref{eq:energy22 of spinor in massive case 1} are given by
	\begin{align}
		\frac{B}{q} & = \frac{2 r^{2} \left(H - 3 L + 2\right)}{H + 5},\label{eq:B/q}\\
		B^2&=  \frac{\left(H - 1\right) \left(L + 1\right)r^2}{2(H + 5)},\label{eq:solution B in massive case 1}\\
		q^2&= \frac{(H+5)\left(H - 1\right) \left(L + 1\right)}{ 8r^2\left(H - 3 L + 2\right)^2 },\label{eq:solution q in massive case 1}\\
		E^2&= - \frac{ \left(H^{2} - 2 H L + 6 H - 4 L + 11\right) \left(H^{2} - 2 H L + 8 H - 16 L + 9\right)}{16 r^2 \left(H - 3 L + 2\right)^2}\label{eq:solution E in massive case 1},
	\end{align}
	where $L=\Lambda r^2$.
	The conditions for the existence of real solutions are $ q^2,B^2,E^2,||\phi^\pm_P||^2_N>0$ and then they hold if and only if $ H,L, $ and $ E $ lie in the range (i), (ii), or (iii) in the following table:
	
	\begin{table}[H]
		\caption{The ranges of $ H, L, $ and $ E $, and the corresponding $N$}
		\centering
		\begin{tabular}{|c|c|c|c|c|}
			\hline
			& Range of $H$ & Range of $L$ & Sign of $E$ & $ N $ \\
			\hline\hline
			(i) & $-8<H<-5$  & $h(H)<L<\infty$ & Negative & $ \widetilde{SL(2,\bR)} $ \\
			\hline
			(ii) & $-5<H<-4$  & $k(H)<L<-1$ & Negative & $ \widetilde{SL(2,\bR)} $ \\
			\hline
			(iii) & $1<H$  & $h(H)<L<k(H)$ & Positive & $ S^3 $ \\
			\hline
		\end{tabular}
	\end{table}
	
	\noindent where $ k(H)=\frac{H+2}{2},\  h(H)=\frac{H^2+8H+9}{2(H+8)}$.
	The details of these calculations are shown in Appendix \ref{sc:Calculations in the section 3}.

	%%%%%%%%%%%%%%%%%%%%%%%%%%%%%%%%%%%%%%%%%%%%%%%%%%%%%%%%%%%%%%%%%%%%%%%%%%%%
	\subsubsection{The case $ (\ell_+,\ell_-)=(1,1)$}
	The equations \eqref{eq:ein01} and \eqref{eq:energy01 of spinor in massive case 2} imply $H=-3$.
	The solutions to the Einstein equations \eqref{eq:ein00},\eqref{eq:ein11},\eqref{eq:ein22} with \eqref{eq:energy00 of spinor in massive case 2},\eqref{eq:energy11 of spinor in massive case 2},\eqref{eq:energy22 of spinor in massive case 2} are given by
	\begin{equation}
		B^2 = - r^2  (\Lambda r^{2} + 1),\ 
		q^2 = - \frac{1}{2r^2(3 \Lambda r^{2} + 1)},\ 
		E^2 = \frac{\Lambda r^{2} + 1}{4r^2(3 \Lambda r^{2} + 1)}\label{eq:solution B,q,E in massive case 2}
	\end{equation}
	The conditions for the existence of real solutions are $ q^2,B^2,E^2,||\phi^\pm_P||^2_N>0$ and then 
	we can verify that it is impossible as follows.
	From $||\phi^\pm_P||^2_N>0$ in \pref{pr:D-M in massive case}, we have $4Er<-4Bq-2$ and $4Er<4Bq-2$, which implies
	\begin{equation}
		2Er<-2|Bq|-1<0,\label{eq:inequality1 in case massive 2}   
	\end{equation}
	especially we have $E<0$.
	% Since we have assumed $E>0$, it is a contradiction.
	
	% Furthermore, in this case, there exists no solutions even if we admit negative $E$.
	% In fact, 
	Since \eqref{eq:solution B,q,E in massive case 2}, $|Bq|=-\sqrt{2}Er$ holds.
	Then \eqref{eq:inequality1 in case massive 2} implies $2Er<2\sqrt{2}Er-1$, so we have $0<E$, a contradiction.
	
	\begin{remark}
		In the case $ H=-5$, there exists probably no solutions, however, it is difficult to show that because of the complexity of the equations.
	\end{remark}

	%%%%%%%%%%%%%%%%%%%%%%%%%%%%%%%%%%%%%%%%%%%%%%%%%%%%%%%%%%%%%%%%%%%%%%%%%%%%
	
	\section{Conclusions and discussions}\label{sc:Conclusions and discussions}
	We considered an E-D-M system consisting of Einstein gravity, an electromagnetic field and two massive chiral spinors coupled to the gauge field.
	We constructed a family of exact solutions to the system.
	The spacetimes are static Sasakian spacetimes, which are direct product manifolds of time and Sasakian space forms.
	The electromagnetic field is a contact magnetic field whose gauge field is proportional to the contact form of the Sasakian space form.
	The two chiral spinors have opposite charges and their total current flows along the Reeb orbits.
	Each chiral spinor consists of two Sasakian quasi-Killing spinors.
	
	We found the solutions whose weak energy condition of the spinors is violated for $N=\widetilde{{\rm SL}(2,\bR)}$.
	Energy conditions are hypotheses that are thought to hold for physically rational matter.
	Interestingly, however, for an arbitrary energy condition, situations violating the energy condition have been reported \cite{Thorne 1988,Borde 1997,Visser 1997} (and references therein).
	Our solutions violating the weak energy condition describe a situation in which the Dirac currents flow uniformly in one direction in an open universe, and these currents generate a global cosmic magnetic field.
	This is not a very strange situation, but it is interesting that the weak energy condition is violated.
	
	In future research, we are interested in constructing solutions to an E-D-M system in non-static stationary spacetimes, expanding generalized R-W models, or Robinson spacetimes.
	We are also interested in using scalar fields rather than spinors.

	%%%%%%%%%%%%%%%%%%%%%%%%%%%%%%%%%%%%%%%%%%%%%%%%%%%%%%%%%%%%%%%%%%%%%%%%%%%%
	\section*{Acknowledgement}
	We consulted H. Ishihara, K. Nakao and H. Yoshino for discussion on the weak energy condition.
	We would like to express our gratitude to them.

	%%%%%%%%%%%%%%%%%%%%%%%%%%%%%%%%%%%%%%%%%%%%%%%%%%%%%%%%%%%%%%%%%%%%%%%%%%%%
	\appendix
	\section{Sasakian frames}\label{sc:sasakian frame}
	In this section, we explain a Sasakian frame.
	According to \cite{Peralta 2021,Nicolaescu 1998}, a Sasakian 3-manifold $(N,\phi,\xi,\eta,g)$ there exists a (local) orthonormal frame $\{e_1=\xi,e_2,e_3\}$ which satisfies
	\begin{align}
		&[\xi,e_2]=-(C_1+1)e_3,\ [\xi,e_3]=(C_1+1)e_2,\\
		&[e_2,e_3]=-2\xi+C_2e_2+C_3e_3,\ C_i\in C^\infty(N),
	\end{align}
	and then the $\phi-$sectional curvature is given by
	\begin{align}
		H=e_2(C_3)-e_3(C_2)-C_2^2-C_3^2+2C_1-1.\label{eq:phi-sec curv}
	\end{align}
	
	On Sasakian space forms of type $S^3,{\rm Nil}_3$ and $\widetilde{SL(2,\mathbb{R})}$, the Lie algebra consisting of Killing vectors are of Bianchi type IX, II and VIII respectively.
	For a suitable basis $\{X_1,X_2,X_3\}$, the structure constants are given by
	\begin{align}
		[X_1,X_2]=-\alpha X_3,\ [X_1,X_3]=\alpha X_2,\ [X_2,X_3]=\beta X_1,
	\end{align}
	where $\alpha=-1,\beta=1$ for type IX, $\alpha=0,\beta=1$ for type II and $\alpha=\beta=-1$ for type VIII.
	
	If a Sasakian manifold has constant $\phi-$sectional curvature $H>-3,H=-3$ and $H<-3$ then it is of type $S^3,{\rm Nil}_3$ and $\widetilde{SL(2,\mathbb{R})}$ respectively.
	Therefore, taking into account \eqref{eq:phi-sec curv}, we obtain a new basis that satisfies \eqref{eq:sasakian frame} by suitable re-scaling of $\{X_i\}$.

	\section{Calculations in section 3}\label{sc:Calculations in the section 3}
	%\subsection{Calculations in the section \ref{subsec:massive case 1}}
	Here we investigate the case of $ (\ell_+,\ell_-)=(1,0)$ in detail.
	The solutions \eqref{eq:solution B in massive case 1},\eqref{eq:solution q in massive case 1}, \eqref{eq:solution E in massive case 1} are obtained by simple algebraic calculation, for example, one easily verify it by computer algebra system.
	We reduce the positivity conditions $ q^2,B^2,E^2,||\phi^\pm_P||^2_N>0$.
	
	We put $ 2\rho_P^2 \coloneqq ||\phi_P^+||_N^2 = ||\phi_P^-||_N^2 $ and investigate the condition $\rho_R^2,\rho_L^2>0$.
	Substituting $ B/q $ and $ Bq = (B/q)q^2 =  \frac{\left(H - 1\right) \left(L + 1\right)}{4 \left(H - 3 L + 2\right)} $ into \eqref{eq:phi_R in massive case 1} and \eqref{eq:phi_L in massive case 1}, we have
	\begin{align}
		\rho_R^2 & = - \frac{2 \left(H - 3 L + 2\right) \left(4 E H r - 12 E L r + 8 E r + H^{2} - 2 H L + 8 H - 16 L + 9\right)}{r \left(H + 5\right) \left(H^{2} - 2 H L + 8 H - 16 L + 9\right)},\\
		\rho_L^2 & = \displaystyle - \frac{2 \left(H - 3 L + 2\right) \left(4 E H r - 12 E L r + 8 E r - H^{2} + 2 H L - 8 H + 16 L - 9\right)}{r \left(H + 5\right) \left(H^{2} - 2 H L + 8 H - 16 L + 9\right)}.
	\end{align}
	Then we obtain
	\begin{align}
		\rho_R^2+\rho_L^2&=\displaystyle - \frac{16 E \left(H - 3 L + 2\right)^{2}}{\left(H + 5\right) \left(H^{2} - 2 H L + 8 H - 16 L + 9\right)},\\
		\frac{\rho_R^2}{\rho_R^2+\rho_L^2}&=\frac{1}{2}+\frac{H^{2} - 2 H L + 8 H - 16 L + 9}{8 E r \left(H - 3 L + 2\right)}>0,\\
		\frac{\rho_L^2}{\rho_R^2+\rho_L^2}&=\frac{1}{2}-\frac{H^{2} - 2 H L + 8 H - 16 L + 9}{8 E r \left(H - 3 L + 2\right)}>0,
	\end{align}
	which implies
	\begin{align}
		&-\frac{1}{2}<\frac{H^{2} - 2 H L + 8 H - 16 L + 9}{8 E r \left(H - 3 L + 2\right)}<\frac{1}{2},\\
		&|H^{2} - 2 H L + 8 H - 16 L + 9| < |4 E r \left(H - 3 L + 2\right)|,\label{eq:inequality 1 in appendix A}
	\end{align}
	and substitute \eqref{eq:solution E in massive case 1} into \eqref{eq:inequality 1 in appendix A}, then
	\begin{align}
		% &(H^{2} - 2 H L + 8 H - 16 L + 9)^2 <- \left(H^{2} - 2 H L + 6 H - 4 L + 11\right) \left(H^{2} - 2 H L + 8 H - 16 L + 9\right)\\
		% &(H^{2} - 2 H L + 8 H - 16 L + 9)^2 + \left(H^{2} - 2 H L + 6 H - 4 L + 11\right) \left(H^{2} - 2 H L + 8 H - 16 L + 9\right)<0\\
		% &(H^{2} - 2 H L + 7 H - 10 L + 10) \left(H^{2} - 2 H L + 8 H - 16 L + 9\right)<0\\
		(H+5)(H+2-2L) \left(H^{2} - 2 (H+8) L + 8 H + 9\right)<0,
	\end{align}
	holds.
	Therefore the conditions to be $ \rho_R^2,\rho_L^2>0$ are given by
	\begin{align}
		&E\left(H + 5\right) \left(H^{2} - 2 H L + 8 H - 16 L + 9\right)<0,\label{eq:condition 1 rho in appendix A}\\
		&(H+5)(H+2-2L) \left(H^{2} - 2 (H+8) L + 8 H + 9\right)<0.\label{eq:condition 2 rho in appendix A}
	\end{align}
	
	We investigate the condition $B^2>0$.
	Since \eqref{eq:solution B in massive case 1}, we have
	\begin{align}
		(H-1)(H+5)(L+1)>0,\label{eq:conditon B in appendix A}
	\end{align}
	and then $q^2>0$ holds.
	
	We investigate the condition $E^2>0$.
	Since \eqref{eq:solution E in massive case 1}, we have
	\begin{align}
		(-2(H+2)L+H^2+6H+11)(-2(H+8)L+H^2+8H+9)<0.\label{eq:condition E in appendix A}
	\end{align}
	
	Therefore the conditions to be $ B^2,q^2,E^2,\rho_R^2,\rho_L^2>0$ are given by \eqref{eq:condition 1 rho in appendix A}, \eqref{eq:condition 2 rho in appendix A}, \eqref{eq:conditon B in appendix A}, \eqref{eq:condition E in appendix A}, and 
	\begin{align}
		H - 3 L + 2\ne0.\label{eq:conditon reg in appendix A}
	\end{align}
	
	We consider only \eqref{eq:condition 2 rho in appendix A}, \eqref{eq:conditon B in appendix A}, \eqref{eq:condition E in appendix A}, \eqref{eq:conditon reg in appendix A} since we can always choose the sign of $E$ so that \eqref{eq:condition 1 rho in appendix A} holds.
	We clarify the possible ranges of $H$ and $L$ for \eqref{eq:condition 2 rho in appendix A}, \eqref{eq:conditon B in appendix A}, \eqref{eq:condition E in appendix A}.
	
	The condition \eqref{eq:condition 2 rho in appendix A} implies the following table.
	%It is easily verified to observe the graphs of $h(H)= \frac{H^{2}  + 8 H + 9}{2(H+8)},\ k(H)=\frac{H+2}{2}$.
	\begin{table}[H]
		\caption{The possible ranges of $H$ and $L$ for \eqref{eq:condition 2 rho in appendix A}}
		\centering
		\begin{tabular}{|c|c|}
			\hline
			Range of $H$ & Range of $L$  \\
			\hline\hline
			$-\infty<H<-8$  & $h<L<k$ or $k<L<h$\\
			\hline
			$H=-8$  & $L<-3$ \\
			\hline
			$-8<H<-5$  & $L<{\rm min}\{k,h\}$ or $ L>{\rm max}\{k,h\}$ \\
			\hline
			$-5<H<\infty$ & $h<L<k$ or $k<L<h$ \\
			\hline
		\end{tabular}
	\end{table}
	\noindent where $h(H)= \frac{H^{2}  + 8 H + 9}{2(H+8)},$ and $ k(H)=\frac{H+2}{2}$. 
	
	The condition \eqref{eq:conditon B in appendix A} implies the following table.
	\begin{table}[H]
		\caption{The possible ranges of $H$ and $L$ for \eqref{eq:conditon B in appendix A}}
		\centering
		\begin{tabular}{|c|c|}
			\hline
			Range of $H$ & Range of $L$  \\
			\hline\hline
			$-\infty<H<-5$  & $-1<L$\\
			\hline
			$-5<H<1$  & $-1>L$ \\
			\hline
			$1<H<\infty$ & $-1<L$ \\
			\hline
		\end{tabular}
	\end{table}
	
	The condition \eqref{eq:condition E in appendix A} implies the following table.
	\begin{table}[H]
		\caption{The possible ranges of $H$ and $L$ for \eqref{eq:condition E in appendix A}}
		\centering
		\begin{tabular}{|c|c|}
			\hline
			Range of $H$ & Range of $L$  \\
			\hline\hline
			$-\infty<H<-8$  & $f<L<h$ or $h<L<f$\\
			\hline
			$H=-8$  & $-\frac{9}{4}>L$ \\
			\hline
			$-8<H<-2$ & $L<{\rm min}\{f,h\}$ or $ L>{\rm max}\{f,h\}$ \\
			\hline
			$H=-2$ & $-\frac{1}{4}<L$ \\
			\hline
			$-2<H<\infty$ & $ f<L<h$ or $ h<L<f$ \\
			\hline
		\end{tabular}
	\end{table}
	\noindent where $ f(H)=\frac{H^2+6H+11}{2(H+2)}$.
	
\begin{figure}[H]
		\centering
		\includegraphics[scale=0.8]{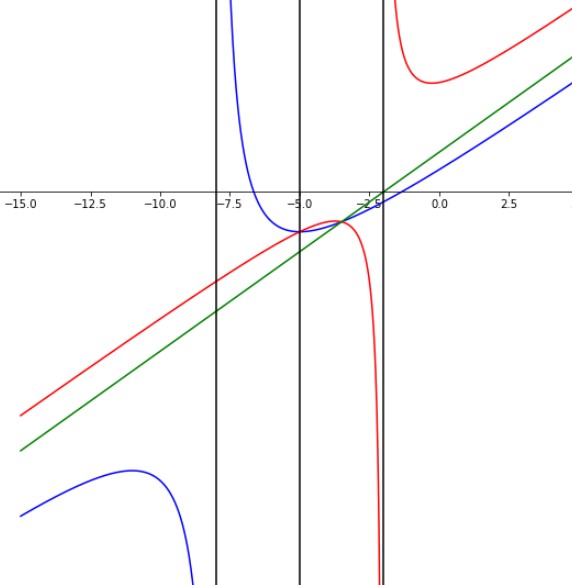}
		\caption{The blue, green and red graphs are the graphs of $ h(H), k(H) $ and $ f(H)$ respectively.  The horizontal axis is  of $ H $.}
		\label{graph}
	\end{figure}

	For each range of $H$, it is easy to decide the possible range of $L$ from the graph \ref{graph}.
	If $ H\in(-\infty,-8)$, it is impossible.
	If $H=-8$, it is impossible.
	If $ H\in(-8,-5)$, $h(H)<L<\infty$ holds.
	If $ H\in(-5,-4)$, $k(H)<L<-1$ holds.
	If $ -4<H<-2$, it is impossible.
	If $H=-2$, it is impossible.
	If $H\in(-2,1)$, it is impossible.
	If $ H\in(1,\infty)$, $h(H)<L<k(H)$ holds.
	Therefore we obtain the following table.
	
	\begin{table}[H]
		\caption{The possible ranges of constants}
		\centering
		\begin{tabular}{|c|c|c|}
			\hline
			Range of $H$ & Range of $L$ & Sign of $E$ \\
			\hline\hline
			$-8<H<-5$  & $h(H)<L<\infty$ & Negative \\
			\hline
			$-5<H<-4$  & $k(H)<L<-1$ & Negative \\
			\hline
			$1<H$  & $h(H)<L<k(H)$ & Positive \\
			\hline
		\end{tabular}
	\end{table}
	
\end{document}